\documentclass[a4paper,10pt]{article}
\usepackage[utf8x]{inputenc}
\usepackage{amssymb,amsfonts,amstext,amsmath,graphicx,epic,amsthm,color,times}

\title{ Stability analysis of Boundary Layer in Poiseuille Flow Through A Modified Orr-Sommerfeld Equation }
\author{A. V. Monwanou\footnote{vincent.monwanou@imsp-uac.org, movins2008@yahoo.fr}, C. H. Miwadinou\footnote
{cl\'ement.Miwadinou@imsp-uac.org} and J. B. Chabi Orou\footnote
{Corresponding author: jean.Chabi@imsp-uac.org, jchabi@yahoo.fr}}
\begin{document}
\maketitle Institut de Math\'ematiques et de Sciences Physiques, BP: 613 Porto Novo, B\'enin

The Abdus Salam International Centre for Theoretical Physics, Trieste, Italy

\begin{abstract}
For applications regarding transition prediction, wing design and control of
boundary layers, the fundamental understanding of disturbance growth in the
flat-plate boundary layer is an important issue. In the present work we investigate the stability of boundary layer in 
Poiseuille flow. We normalize pressure and time by inertial and viscous effects. 
The disturbances are taken to be periodic in the spanwise
 direction and time. We present a set of linear governing equations for the parabolic evolution of
wavelike disturbances. Then, we derive modified Orr-Sommerfeld equations that can be applied in the layer.
Contrary to what one might think, we find that Squire's theorem is not applicable for the boundary layer. We find also that normalization
 by inertial or viscous effects leads 
to the same order of stability or instability. For the $2D$ disturbances flow ($\theta=0$), we found the same critical Reynolds 
number for our two normalizations. This value coincides with the one we know  for neutral stability of the known Orr-Sommerfeld 
equation. We noticed also that for all overs values of $k$ in the case $\theta=0$ correspond the same values of $Re_\delta$ at $c_i=0$ 
whatever the normalization. We therefore conclude that in the boundary layer with a 2D-disturbance, we have the same neutral stability
 curve whatever the normalization. We find also that for a flow
with hight hydrodynamic Reynolds number, the neu-
tral disturbances in the boundary layer are two-dimensional.
At last, we find that transition 
from stability to instability or the opposite
 can occur according to the Reynolds number and the wave number.

\end{abstract}

\section{Introduction}
 Boundary-layer theory is crucial in understanding why certain phenomena occur. It is well known that the instability of
 boundary layers is sensitive to the mean velocity profile, so that a small distortion to the basic flow may have a
 detrimental effect on its stability. Prandtl (1904)\textbf{\cite{Lan}} proposed that viscous effects would be confined to thin 
layers adjacent 
to boundaries in the case of the motion of fluids with very little viscosity i.e. in the case of flows for which the
 characteristic 
Reynolds number, $R_e$, is large. In a more general sense we will use boundary-layer theory (BLT) to refer to any 
large-Reynolds-number. 
Ho and Denn studied low Reynolds number stability for plane Poiseuille flow by using
a numerical scheme based on the shooting method. They found that at low Reynolds numbers no instabilities occur, but the 
numerical
 method led to artificial instabilities.Lee and Finlayson used a similar numerical method to study both Poiseuille and
Couette flow, and confirmed the absence of instabilities at low Reynolds number.
 R. N. RAY et al \textbf{\cite{ray}} investigated the linear stability of plane Poiseuille flow at small Reynolds number
 of a conducting Oldroyd fluid
 in the presence of magnetic field. They found that viscoelastic parameters have destabilizing effect and magnetic field has a 
stabilizing effect in the field of flow but no instabilities are found. 

   In this paper, we study the linear stability of boundary layer in a  plane\\ Poiseuille flow. For this, we derive two 
fourth-order
 equations name modified fourth-order
Orr-Sommerfeld equations governing the stability analysis in boundary layer for the flow. The first is obtained by making 
dimensionless
 quantities by the inertial effects. The second takes into account the form adopted by the rheologists i.e. make the quantities
 dimensionless by normalizing by the viscous effects. This allowed us to see the effect of each type of normalization on the 
stability
 in the boundary layer. So, we solve numerically the corresponding eigenvalues problems. We employ Matlab in all our numerical
 computations to find eigenvalues.

The paper is organized as follows. In the second section the boundary layer theory is presented. In the third section we present
 the general formulation, highlighting the fundamental equations that model the flat-plate boundary layer flow according to the 
normalization
 by inertial and viscous  effects. In the fourth section the modified Orr-Sommerfeld equations governing the stability analysis
 in boundary layer are checked and in the fifth section, analysis of the stability is investigated. The conclusions and
 perspectives
are presented in the final section.
\section{Boundary Layer Theory}
When applying the theory of complex potential around an airfoil in considering the model of inviscid incompressible irrotational 
plan flow, 
we know that the model allows to deduce the lift but the drag is zero. This is contrary to experimental observations which show 
that
 the drag affects all flows of real fluids. These are viscous. They adhere to the walls and the tangential component of 
the velocity
 is zero if the wall is fixed. The latter condition can be satisfied by the perfect fluid. Moreover, the irrotational condition 
 is far from reality as we know that the production of vorticity occurs at the walls. To remedy the deficiencies of the theory of
 perfect fluid, it must appeal to the theory of the boundary layer which is a necessary correction for flows with high Reynolds
 numbers. Theoris the boundary layer is due to L. Prandtl\textbf{\cite{Lan}}. The boundary layer is the area of ​​the flow 
which is close to the wall or 
of an obstacle present in a uniform flow at the upstream infinity or on the confining walls of internal flow. Within the boundary
 layer is a thin zone, it is estimated that viscous effects are of the same magnitude as the inertial effects. The boundary layer
 is the place of intense generation of vorticity which will not diffuse into the area outside thereof. This leads to a very modern
 concept of comprehensive approach to the problem by breaking it down into two areas: firstly the boundary layer where we will
 consider the viscous effects in a model of simplified Navier-Stokes and other from the outer area where we will use the complex
 potential theory in the inviscid incompressible flow. This outer zone has speeds which are of the same order of magnitude as
that of the incident flow.

The boundary layer along an obstacle is therefore thin since the fluid travel great distances downstream of the leading edge
 during the time interval during which the vortex diffuse only a small distance from the wall. The creation of vorticity in the
 boundary layer allows the physical realization of the fluid flow around the profile. This movement gives rise to a wake in the
 area near the trailing edge. The importance of the wake depend on the shape of the obstacle and the angle of incidence of the 
upstream flow at the leading edge.

We consider incompressible flow of a fluid with constant density $\rho$ and dynamic viscosity $\mu$, past a body with typical 
length $L$. We assume that a typical velocity scale is $U$, and the Reynolds number is given by

\begin{equation}
 R_e=\frac{\rho U L}{\mu}\gg 1.
\end{equation}

For simplicity we will, for the most part, consider two-dimensional incompressible flows, although many of our statements can be 
generalised to three-dimensional flowsand/or compressible flows.

Boundary Layer Theory applies to flows where there are extensive inviscid regions separated by thin shear layers, say, of typical 
width $\delta\ll L$. For one such shear layer take local dimensional Cartesian coordinates $\hat{x}$ and $\hat{y}$ along 
and across the 
shear layer respectively. Denote the corresponding velocity components by $\hat{u}$ and $\hat{v}$ respectively, pressure 
by $\hat{p}$ 
and time by $\hat{t}$. On the basis of scaling arguments it then follows that

\begin{equation}
 \delta\sim R_e^{-\frac{1}{2}} L \ll L.
\end{equation}

Further, it can also be deduced that the key approximations in classical Boundary Layer Theory are that the pressure is
 constant across 
the shear layer i.e.

\begin{equation}
 0=-\hat{p}_{\hat{y}},
\end{equation}
and that streamwise diffusion is negligible, i.e. if  $\bullet$  represents any variable 

\begin{equation}
 \bullet_{\hat{y}\hat{y}}\gg \bullet_{\hat{x}\hat{x}}
\end{equation}
 
The former approximation is more significant dynamically.

Now, using the transformations

\begin{equation}
 (\hat{x}, \hat{y}, \hat{t}, \hat{u}, \hat{v}, \hat{p})\rightarrow (L x, R_e^{-\frac{1}{2}} L y, U^{-1} L t, Uu, R_e^{-\frac{1}{2}} U v, \rho U^2 p )
\end{equation}

and taking the limit $R_e\rightarrow \infty $, the Boundary Layer Theory equations can be deduced from the Navier-Stokes equations:

\begin{equation}
 u_t+uu_x+vu_y=-p_x+u_{yy},
\end{equation}

\begin{equation}
 0=-p_y,\;\;\;\;\; u_x+v_y=0
\end{equation}

 For flow past a rigid body the appropriate boundary conditions are
\begin{equation}
 u=v=0 \;\;\; on \;\;\; y=0\;\;\; and\;\;\; u\rightarrow U(x,t)\;\;\; as\;\;\; y\rightarrow \infty,
\end{equation}

where $U(x,t)$ is the inviscid slip velocity past the body. Further, from (6) evaluated at the edge of the boundary layer

\begin{equation}
 -p_x=U_t+UU_x.
\end{equation}

We define the viscous blowing velocity out of the boundary layer to be 

\begin{equation}
 v_b(x,t)=\lim_{y\rightarrow \infty}(v+U_x(x,t)y).
\end{equation}

$v_b$ indicates the strength of blowing, or suction, at the edge of the boundary layer induced by viscous effects. It is good
 diagnostic 
for dynamically significant effects within the boundary layer much better than, say, the wall shear $u_y(x,0,t)$ which can remain 
regular while $v_b(x,t)$ becomes unbounded. 

\section{General formulation}
Consider an incompressible boundary layer over a flat plate as illustrated
in the following Figure\textbf{\cite{hen}}.

\begin{figure}[htbp]
 \begin{center}
  \includegraphics[width=12cm]{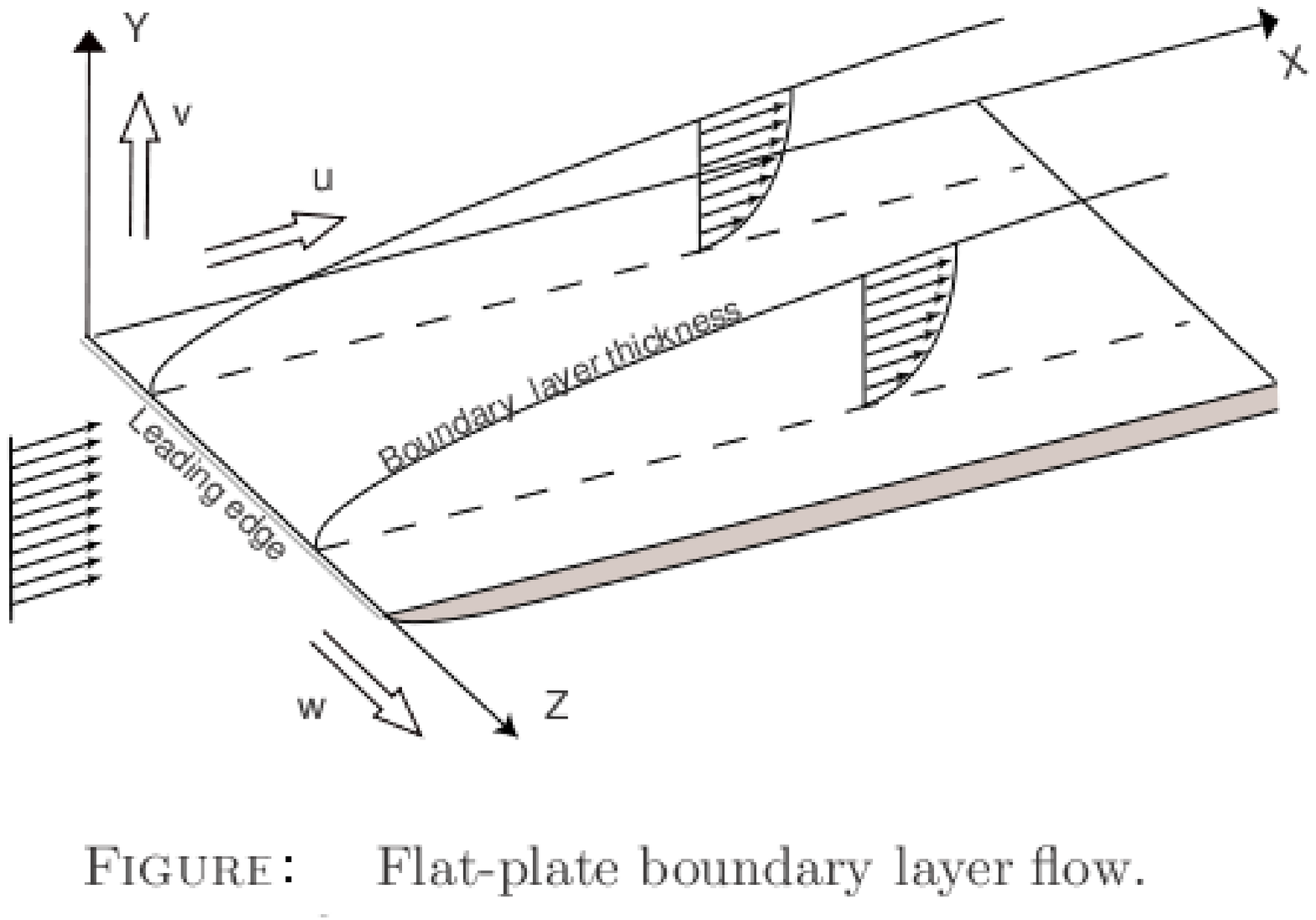}
  \end{center}
 \end{figure}

 The streamwise coordinate $x$ is scaled
with the length scale $l$, which is a fixed distance from the leading edge. The
wall-normal and spanwise coordinates $y$ and $z$, respectively, are scaled with the
boundary-layer parameter $\delta=\sqrt{\nu l/U_{\infty}}$ , where $\nu$ is the kinematic viscosity of
the fluid and $U_{\infty}$ is the streamwise freestream velocity at the distance $l$ from
the leading edge. The streamwise velocity $u$ is scaled with $U_{\infty}$ , while the wall-
normal and spanwise velocities $v$ and $w$, respectively, are scaled with $U_{\infty} \delta /l$.
The pressure $p$ is scaled with $\rho U^2_{\infty}\delta^2/l^2$ and the time $t$ is scaled with $l/U_{\infty}$ . The
Reynolds numbers used here are defined as $Re_l=U_{\infty}l/\nu$ and $Re_{\delta}=U_{\infty}\delta/\nu$.
It is useful to note the relations $l/\delta=Re_{\delta}=\sqrt{Re_l}$ .
We want to study the linear stability of a high Reynolds number flow.
The non-dimensional Navier–Stokes equations for an incompressible flow where pressure and time are normalized by inertial effects

\begin{equation}
\frac{\partial \bar{u}}{\partial x}  +\frac{\partial \bar{v}}{\partial y}+\frac{\partial \bar{w}}{\partial z}=0,
\end{equation}

\begin{equation}
 \frac{D\bar{u}}{Dt}=-\frac{1}{Re^2_\delta}\frac{\partial \bar{p}}{\partial x}+\frac{1}{Re^2_\delta}\frac{\partial^2 \bar{u}}{\partial x^2}+\frac{\partial^2 \bar{u}}{\partial y^2}+\frac{\partial^2 \bar{u}}{\partial z^2},
\end{equation}

\begin{equation}
 \frac{D\bar{v}}{Dt}=-\frac{\partial \bar{p}}{\partial y}+\frac{1}{Re^2_\delta}\frac{\partial^2 \bar{v}}{\partial x^2}+\frac{\partial^2 \bar{v}}{\partial y^2}+\frac{\partial^2 \bar{v}}{\partial z^2},
\end{equation}

\begin{equation}
 \frac{D\bar{w}}{Dt}=-\frac{\partial \bar{p}}{\partial z}+\frac{1}{Re^2_\delta}\frac{\partial^2 \bar{w}}{\partial x^2}+\frac{\partial^2 \bar{w}}{\partial y^2}+\frac{\partial^2 \bar{w}}{\partial z^2},
\end{equation}

where $\frac{D}{Dt}=\frac{\partial}{\partial t}+\bar{u} \frac{\partial}{\partial x}+\bar{v}\frac{\partial}{\partial y}+\bar{w}\frac{\partial}{\partial z}$

are linearized around a two-dimensional, steady base flow $(U (x, y), V (x, y), 0)$ to obtain the stability equations for the 
spatial evolution of three-dimensional,
time-dependent disturbances $(u(x, y, z, t), v(x, y, z, t), w(x, y, z, t), p(x, y, z, t))$ .
The base flow and the disturbances are scaled in the same way. The disturbances are taken to be periodic in the spanwise
 direction and time, which allows
us to assume solutions of the form

\begin{equation}
 f=\hat{f}(x,y)exp(iRe_\delta \int_{x_0}^x \alpha(x)dx+i\beta z-i\omega t)
\end{equation}

where $f$ represents either one of the disturbances $u$, $v$, $w$ or $p$. The complex
streamwise wave number $\alpha$ captures the fast wavelike variation of the modes
and is therefore scaled with $1/\delta$. $\alpha$ itself is assumed to vary slowly with $x$.
Since $x$ is scaled with $l$, the factor $Re_\delta$ appears in front of the integral. The $x$-dependence in the amplitude
 function $\hat{f}$ includes the weak variation of the
disturbances. The real spanwise wave number $\beta$ and angular frequency $\omega$ are
scaled in a consistent way with $z$ and $t$, respectively. Introducing the assump-
tion (11) in the linearized Navier–Stokes equations and neglecting all third order
terms in $1/Re_\delta$ or higher, we arrive at the parabolized stability equations in boundary-layer scalings

\begin{equation}
 \hat{u}_x+iRe_\delta \alpha \hat{u}+\hat{v}_y+i\beta\hat{w}=0,
\end{equation}

\begin{equation}
 (U_x+iRe_\delta \alpha U-i\omega)\hat{u}+U\hat{u}_x+V\hat{u}_y+U_y\hat{v}+\frac{\hat{p}_x}{Re^2_\delta}+\frac{i\alpha\hat{p}}{Re_\delta}=\hat{u}_{yy}-k^2\hat{u},
\end{equation}

\begin{equation}
 (V_y+iRe_\delta \alpha U-i\omega)\hat{v}+U\hat{v}_x+V_x\hat{u}_y+V\hat{v}_y+\hat{p}_y=\hat{v}_{yy}-k^2\hat{v},
\end{equation}

\begin{equation}
 (iRe_\delta \alpha U-i\omega)\hat{w}+U\hat{w}_x+V\hat{w}_y+i\beta\hat{p}=\hat{w}_{yy}-k^2\hat{w},
\end{equation}

where $k^2=\alpha^2+\beta^2$.

If we normalize pressure and time by viscous effects i.e. $t$ scaled with $\rho l^2/\mu$ and $p$ scaled with $l/\mu U_\infty$ the
 Navier-Stokes equations take the following forms

\begin{equation}
\frac{\partial \bar{u}}{\partial x}  +\frac{\partial \bar{v}}{\partial y}+\frac{\partial \bar{w}}{\partial z}=0,
\end{equation}

\begin{equation}
 \frac{1}{Re^2_\delta} \frac{\partial \hat{u}}{\partial t}+\hat{u}\frac{\partial \bar{u}}{\partial x}  +\hat{v}\frac{\partial \bar{u}}{\partial y}+\hat{w}\frac{\partial \bar{u}}{\partial z}=-\frac{1}{Re^2_\delta}\frac{\partial \bar{p}}{\partial x}+\frac{1}{Re^2_\delta}\frac{\partial^2 \bar{u}}{\partial x^2}+\frac{\partial^2 \bar{u}}{\partial y^2}+\frac{\partial^2 \bar{u}}{\partial z^2},
\end{equation}

\begin{equation}
 \frac{1}{Re^2_\delta} \frac{\partial \hat{v}}{\partial t}+\hat{u}\frac{\partial \bar{v}}{\partial x}  +\hat{v}\frac{\partial \bar{v}}{\partial y}+\hat{w}\frac{\partial \bar{v}}{\partial z}=-\frac{\partial \bar{p}}{\partial y}+\frac{1}{Re^2_\delta}\frac{\partial^2 \bar{v}}{\partial x^2}+\frac{\partial^2 \bar{v}}{\partial y^2}+\frac{\partial^2 \bar{v}}{\partial z^2},
\end{equation}

\begin{equation}
  \frac{1}{Re^2_\delta} \frac{\partial \hat{w}}{\partial t}+\hat{u}\frac{\partial \bar{w}}{\partial x}  +\hat{v}\frac{\partial \bar{w}}{\partial y}+\hat{w}\frac{\partial \bar{w}}{\partial z}=-\frac{\partial \bar{p}}{\partial z}+\frac{1}{Re^2_\delta}\frac{\partial^2 \bar{w}}{\partial x^2}+\frac{\partial^2 \bar{w}}{\partial y^2}+\frac{\partial^2 \bar{w}}{\partial z^2}.
\end{equation}

The linearized Navier-Stokes equations with the previous disturbances under the same considerations become
\begin{equation}
 \hat{u}_x+iRe_\delta \alpha \hat{u}+\hat{v}_y+i\beta\hat{w}=0,
\end{equation}

\begin{equation}
 (U_x+iRe_\delta \alpha U-\frac{i \omega}{Re^2_\delta})\hat{u}+U\hat{u}_x+V\hat{u}_y+U_y\hat{v}+\frac{\hat{p}_x}{Re^2_\delta}+\frac{i\alpha\hat{p}}{Re_\delta}=\hat{u}_{yy}-k^2\hat{u},
\end{equation}

\begin{equation}
 (V_y+iRe_\delta \alpha U-\frac{i \omega}{Re^2_\delta})\hat{v}+U\hat{v}_x+V_x\hat{u}_y+V\hat{v}_y+\hat{p}_y=\hat{v}_{yy}-k^2\hat{v},
\end{equation}

\begin{equation}
 (iRe_\delta \alpha U-\frac{i \omega}{Re^2_\delta})\hat{w}+U\hat{w}_x+V\hat{w}_y+i\beta\hat{p}=\hat{w}_{yy}-k^2\hat{w},
\end{equation}

where $k^2=\alpha^2+\beta^2$.

\section{Modified Orr-Sommerfeld Equation}
Considering temporal the stability problem with $\omega=\alpha c$ and $(\alpha,\beta)$ real, we will simplify the problem.
 Our strategy will be first to eliminate $\hat{u}$, $\hat{w}$, $\hat{p}$ to leave a single equation in $\hat{v}$. This can be
 used finally to determine
 the linear stability (or instability) in the boundary layer oy the base flow. Remember that $c=c_r+ic_i$ and if $c_i<0$, the flow 
is stable; $c_i>0$, the flow is unstable and we have neutral stability if $c_i=0$.

Taking $i\alpha Re_\delta (17)$ +$i\beta (19)$ we get 

\begin{eqnarray}
 i \alpha Re_\delta U (i\alpha Re_\delta \hat{u}+i\beta \hat{w})-i\omega (i\alpha Re_\delta \hat{u}+i\beta \hat{w})+
U(i \alpha Re_\delta \hat{u}_x+i \beta \hat{w}_x)+ \nonumber \\
+i\alpha Re_\delta U_y \hat{v}-\beta^2 \hat{p}+\frac{i\alpha \hat{p}_x}{Re_\delta}
-\alpha^2 \hat{p}=(\frac{d^2}{dy^2}-k^2)(i\alpha Re_\delta \hat{u}+i\beta \hat{w}).
\end{eqnarray}

Using the continuity equation (16), (28) becomes
\begin{eqnarray}
  i \alpha Re_\delta U (\hat{u}_x+\hat{v}_y)-i\omega (\hat{u}_x+\hat{v}_y)+
U \frac{d}{dx}(\hat{u}_x+\hat{v}_y)+ \nonumber \\
-i\alpha Re_\delta U_y \hat{v}+k^2 \hat{p}-\frac{i\alpha \hat{p}_x}{Re_\delta}
=(\frac{d^2}{dy^2}-k^2)(\hat{u}_x+\hat{v}_y).
\end{eqnarray}

Operating across (29) with  $d/dy$, using the assumptions in boundary layer and the equation (18) we get the modified 
Orr-Sommerfeld equation for the boundary layer

\begin{equation}
 (i\alpha U Re_\delta-ic\alpha)(\frac{d^2}{dy^2}-k^2)\hat{v}-i\alpha Re_\delta U^{''}\hat{v}=(\frac{d^2}{dy^2}-k^2)^2\hat{v}.
\end{equation}

Considering normalization with viscous effects we would have

\begin{equation}
 (i\alpha U Re_\delta-\frac{ic\alpha}{Re^2_\delta})(\frac{d^2}{dy^2}-k^2)\hat{v}-i\alpha Re_\delta U^{''}\hat{v}=(\frac{d^2}{dy^2}-k^2)^2\hat{v}.
\end{equation}
\section{Stability Analysis}
We put equations (30)-(31) respectively in the form 
\begin{equation}\label{iner}
 (Re_\delta U (\frac{d^2}{dy^2}-k^2)-Re_\delta U^{''}+i \alpha^{-1}(\frac{d^2}{dy^2}-k^2))\hat{v}=c(\frac{d^2}{dy^2}-k^2)\hat{v}
\end{equation}
and
\begin{equation}\label{vis}
 (Re^3_\delta U (\frac{d^2}{dy^2}-k^2)-Re^3_\delta U^{''}+i \alpha^{-1}Re^2(\frac{d^2}{dy^2}-k^2))\hat{v}=c(\frac{d^2}{dy^2}-k^2)\hat{v}.
\end{equation}
In order to investigate the application of the Squire theorem, consider first the normalization with inertial effects (\ref{iner}). Often, 
however, we are only interested in the instability that appears first as the control parameter $Re_{\delta}$ is increased. In this case, 
Squire's theorem tells us that we need only consider $2D$ disturbances.

 Consider a base state $U(y)$. Imagine a growing $3D$ disturbance to this base state at Reynolds number $Re_{\delta 3 D}$, with wavenumbers 
 $\alpha_{3D}$, $\beta_{3D}$, and $k_{3D}^2=\alpha_{3D}^2+\beta_{3D}^2$. This corresponds to a solution $c$, $\hat{v}$ with $c_i>0$( of 
the modified Orr-Sommerfeld equation 
\begin{equation}\label{td}
  (i\alpha_{3D} U Re_{\delta 3 D}-ic\alpha_{3D})(\frac{d^2}{dy^2}-k_{3D}^2)\hat{v}-i\alpha_{3D} Re_{\delta 3 D} U^{''}\hat{v}=(\frac{d^2}{dy^2}-k_{3D}^2)^2\hat{v}.
\end{equation}
 Now consider a $2D$ disturbance at a Reynolds number $Re_{\delta 2D}$. This has $\beta_{2D}=0$, $k_{2D}=\alpha_{2D}$ and must satisfy 
the $2D$ modified Orr-Sommerfeld equation
\begin{equation}
 (i\alpha_{2D} U Re_{\delta 2D}-ic\alpha_{2D})(\frac{d^2}{dy^2}-\alpha_{2D}^2)\hat{v}-i\alpha_{2D} Re_{\delta 2D} U^{''}\hat{v}=(\frac{d^2}{dy^2}-\alpha_{2D}^2)^2\hat{v}.
\end{equation}
For values $\alpha_{3D} Re_{\delta 3 D} =\alpha_{2D}  Re_{\delta 2D}$, $\alpha_{3D}=\alpha_{2D}$ and $k_{3D}=\alpha_{2D}$, this $2D$ 
modified Orr-Sommerfeld equation has the form
\begin{equation}
 (i\alpha_{3D} U Re_{\delta 3D}-ic\alpha_{3D})(\frac{d^2}{dy^2}-\alpha_{2D}^2)\hat{v}-ik_{3D} Re_{\delta 3D} U^{''}\hat{v}=(\frac{d^2}{dy^2}-k_{3D}^2)^2\hat{v},
\end{equation}
which is exactly the same as \ref{td}. It must therefore have the same growing solution $c, \hat{v}$ with $c_i>0$.

Therefore, corresponding to the growing $3D$ disturbance at $Re_{\delta 3D}$ with  $\alpha_{3D}$, $\beta_{3D}$, and $k_{3D}^2=\alpha_{3D}^2+\beta_{3D}^2$, 
there exists a growing $2D$ disturbance at $Re_{\delta 2D}$ with $k_{3D}=\alpha_{2D}$ and $\alpha_{2D}=\alpha_{3D}$. These conditions lead 
to $k_{3D}=\alpha_{3D}$ and we get that the disturbances are two-dimensional. Finally, we get $Re_{\delta 2D}=Re_{\delta 3D}$, not 
$Re_{\delta 2D}\leq Re_{\delta 3D}$. And so, we can not applied the Squire's theorem in the boundary layer of the flow.

Using the same assumptions in (\ref{vis}), we get first $Re_{\delta 2D}^2=\frac{\alpha_{2D}}{\alpha_{3D}}Re_{\delta 3D}^2$ with $k_{2D}=k_{3D}$ i.e. 
$\alpha_{2D}=k_{3D}$ and secondly $Re_{\delta 2D}=\frac{\alpha_{3D}}{\alpha_{2D}}Re_{\delta 3D}^2 $ with $k_{2D}=k_{3D}$ i.e. 
$\alpha_{2D}=k_{3D}$. So we have first $Re_{\delta 2D}\geq Re_{\delta 3D}$ and secondly $Re_{\delta 2D}\leq Re_{\delta 3D}$  because 
$k_{3D}\geq \alpha_{3D}$. We see therefore that we must take $Re_{\delta 2D} = Re_{\delta 3D}$ and so $k_{3D}=\alpha_{3D}$. We find the 
same result that the disturbances are two-dimensional and then Squire's theorem can't be applied.
 
Finally, we still consider our three-dimensional disturbances so without using the theorem of Squire. Thus we write $\alpha=k cos\theta$
 with $\theta=(\vec{k}_x,\vec{k})$. This will allow us to deduce numerically whether the application of Squire's theorem in the boundary layer. 
Indeed $\theta=0$ corresponds to a two dimensional disturbance i.e $k=\alpha$.

We employ Matlab (Windows Version) in all our numerical computations to find
eigenvalues. 
A Poiseuille flow with the basic profile 
\begin{equation}
 U(y)=1-y^2
\end{equation}
is considered.

The eigenvalue problems (32)-(33) are solved numerically with the suitable boundary conditions. The solutions are found in a 
layer bounded at $z=\pm 1$ with $U(\pm1)=0$. 
The results of calculations are presented in the following figures. In each group of figures, the first one is figure $a)$, the second
is figure $b)$ and the third is figure $c)$.
First we present the figures relatives to the eigenvalue problem (32).

For a fixed $k=1$, we get figure 1 of $c_i$ vs $R$ for sequential values of $\theta$ in which $a)$ shows the entire graph. $b)$ 
and
 $c)$ are the magnified versions of $a)$.

For a fixed $k=2$, we get figure 2 of $c_i$ vs $R$ for sequential values of $\theta$ in which $a)$ shows the entire graph. $b)$ 
and
 $c)$ are the magnified versions of $a)$.

For a fixed $k=4$, we get figure 3 of $c_i$ vs $R$ for sequential values of $\theta$ in which $a)$ shows the entire graph. $b)$ 
and
 $c)$ are the magnified versions of $a)$.


Secondly we present the figures relatives to the eigenvalue problem (33).

For a fixed $k=1$, we get figure 4 of $c_i$ vs $R$ for sequential values of $\theta$ in which $a)$ shows the entire graph. $b)$ and
 $c)$ are the magnified versions of $a)$.

For a fixed $k=2$, we get figure 5 of $c_i$ vs $R$ for sequential values of $\theta$ in which $a)$ shows the entire graph. $b)$ and
 $c)$ are the magnified versions of $a)$.


Through the figures $1$ and $4$, it is easy to see that if we take a curve with $\theta\neq0$ in the instability area $(c_i>0)$, 
we don't have necessary in the two normalization  $Re_{\delta}(\theta=0)\leq Re_{\delta}(\theta\neq 0)$.  We also see through the figures that
 the normalization of time and pressure by inertial effects or viscous effects lead to the same order of stability/instability in the
 boundary layer. This confirms that in the boundary layer, viscous forces are on the same
 magnitude as the inertial forces i.e. the local Reynolds number is on unity order. We see also through the figures that at low
 Reynolds number the flow is stable but if the Reynolds number increase, unstability appears. The increase of the wave number
 induce also the stability of the flow. By figures $6$ and $7$ we find the first value of $Re_\delta$ and $\alpha$ for which the first 
transition from stability to instability occurs i.e. $c_i=0$ in the case $\theta=0$. We find the values $(Re_\delta,\alpha)=(5772, 1.02)$ 
which corresponds exactly  to the one we know as critical value of the neutral stability of Poiseuille flow. Note that this value is
 the same in figure $6-a)$ which corresponds to normalization by inertial effects and in figure $6-b)$ which corresponds to normalization by 
viscous effects. We noticed also that for all overs values of $k$ in the case $\theta=0$ correspond the same values of $Re_\delta$ at $c_i=0$ 
whatever the normalization. We therefore conclude that in the boundary layer with a 2D-disturbance, we have the same neutral stability
 curve whatever the normalization.
\begin{figure}[htbp]
 \begin{center}
  \includegraphics[width=10cm]{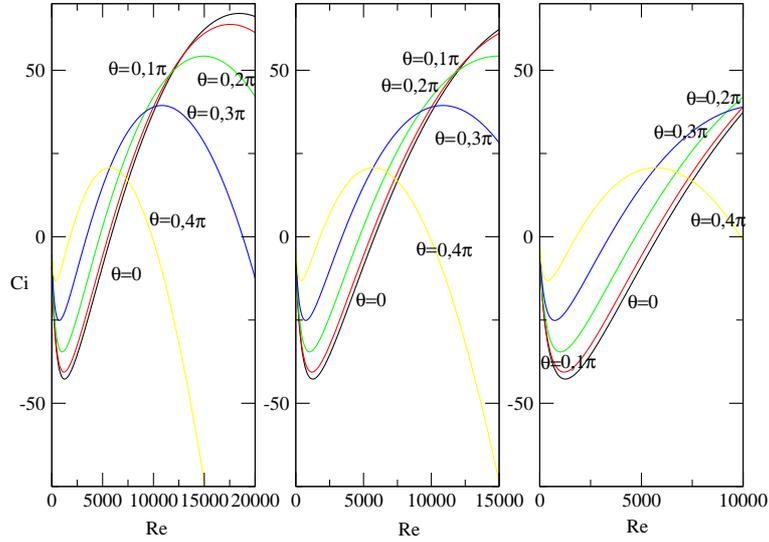}
  \end{center}
  \caption{(a) Growth rate $C_i$ vs. Reynolds number $Re$ for $k=1$; (b) zoom of (a) to small values of k; (c) zoom of (b) to small
 values of $Re$. }
 \end{figure}
\begin{figure}[htbp]
 \begin{center}
  \includegraphics[width=10cm]{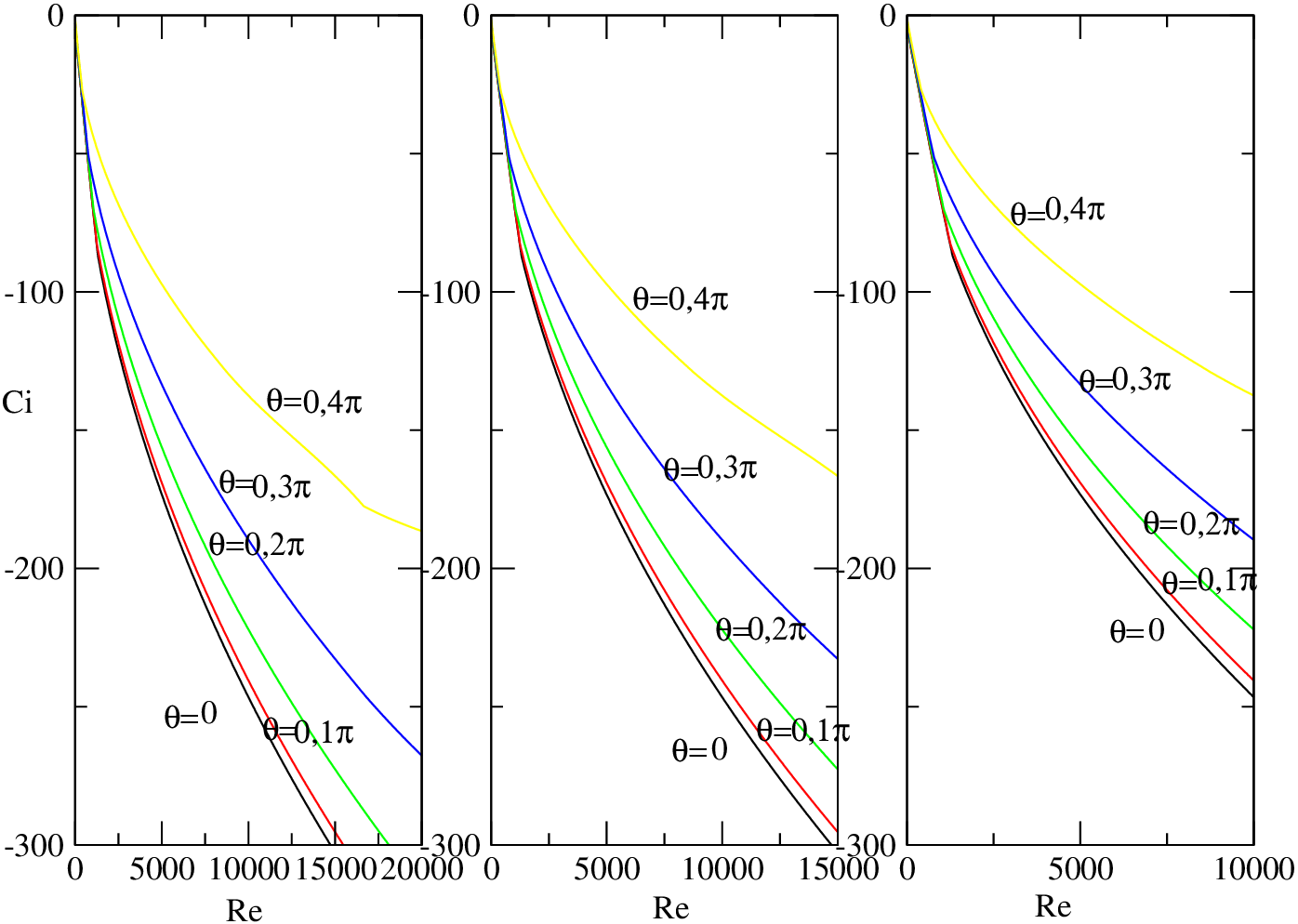}
  \end{center}
  \caption{(a) Growth rate $C_i$ vs. Reynolds number $Re$ for $k=2$; (b) zoom of (a) to small values of k; (c) zoom of (b) to small
 values of $Re$. }
 \end{figure}
\begin{figure}[htbp]
 \begin{center}
  \includegraphics[width=10cm]{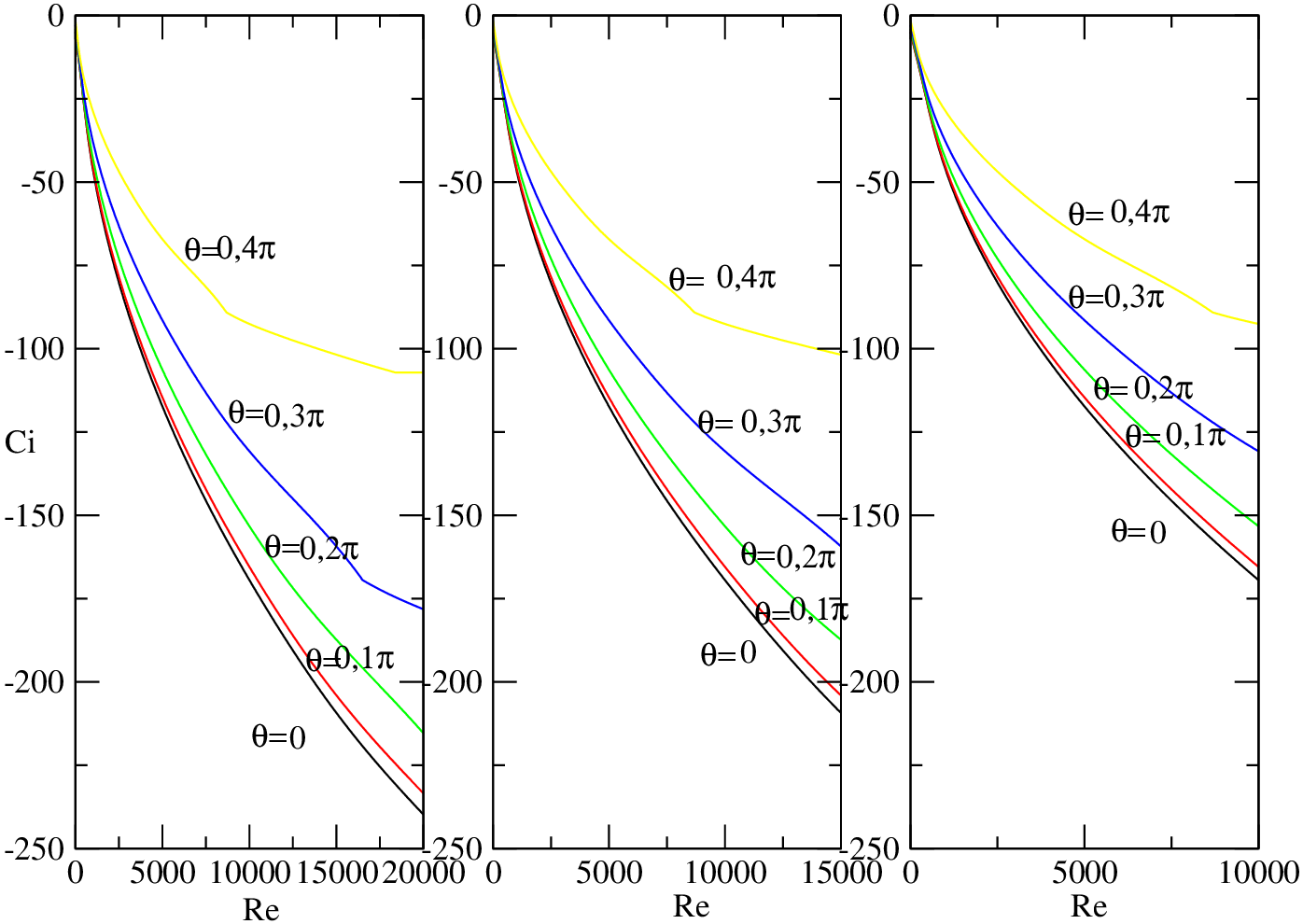}
  \end{center}
 \caption{(a) Growth rate $C_i$ vs. Reynolds number $Re$ for $k=4$; (b) zoom of (a) to small values of k; (c) zoom of (b) to small
 values of $Re$. }
 \end{figure}
\begin{figure}[htbp]
 \begin{center}
  \includegraphics[width=10cm]{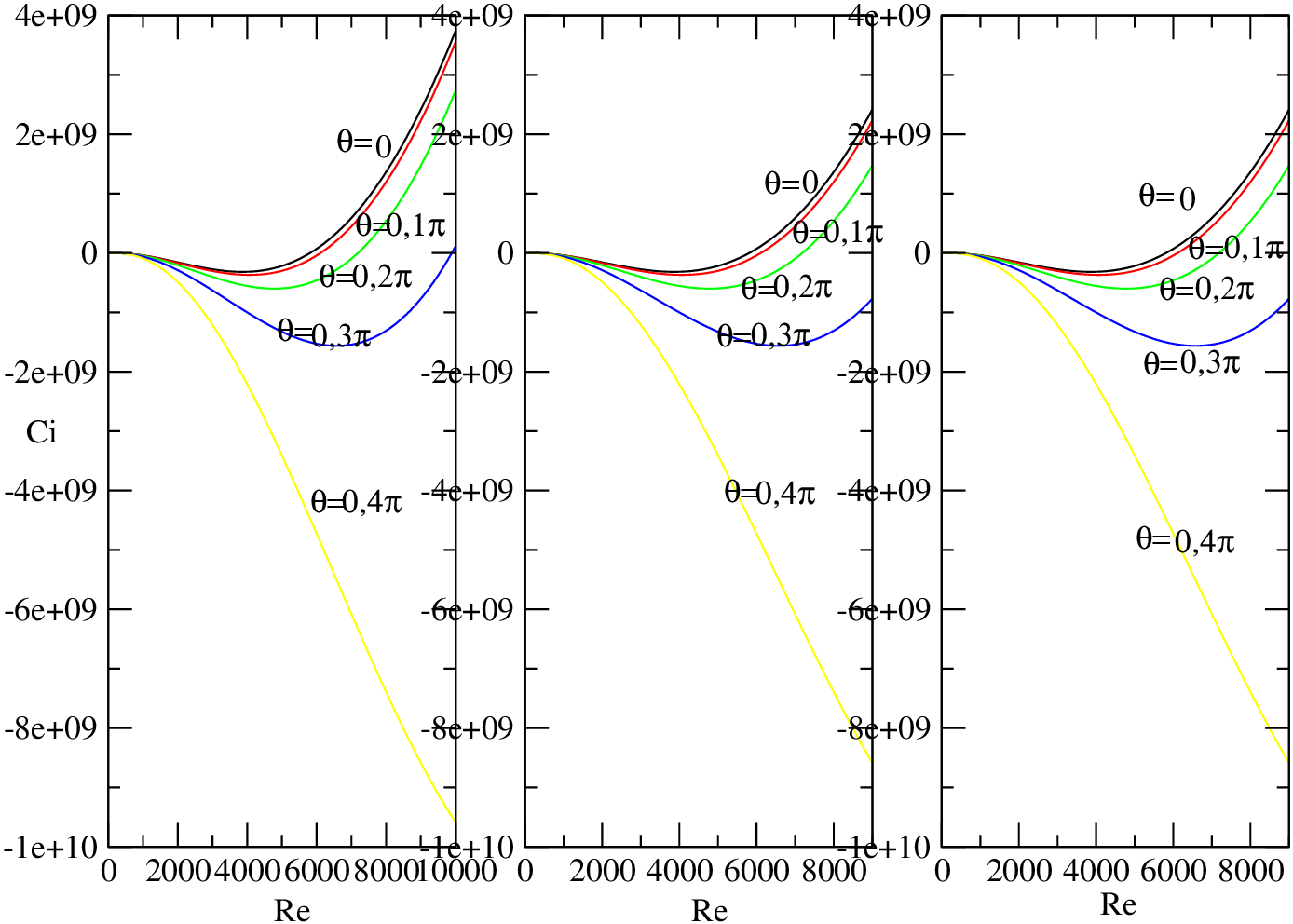}
  \end{center}
  \caption{(a) Growth rate $C_i$ vs. Reynolds number $Re$ for $k=1$; (b) zoom of (a) to small values of k; (c) zoom of (b) to small
 values of $Re$. }
 \end{figure}

\begin{figure}[htbp]
 \begin{center}
  \includegraphics[width=10cm]{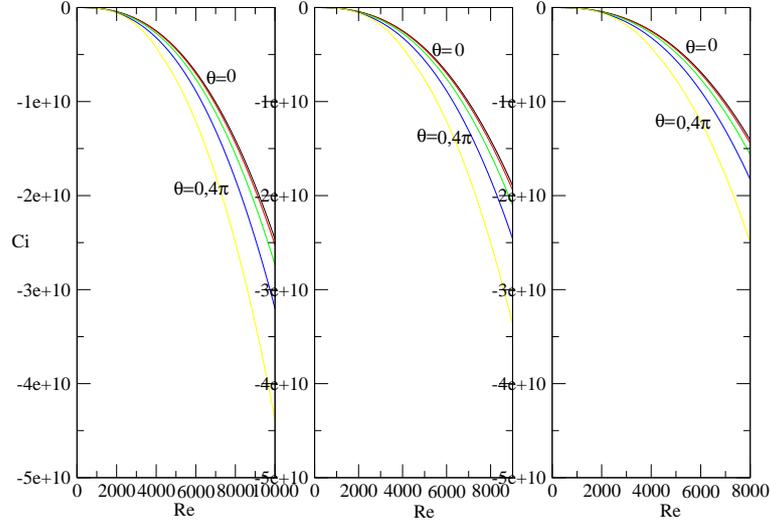}
  \end{center}
  \caption{(a) Growth rate $C_i$ vs. Reynolds number $Re$ for $k=2$; (b) zoom of (a) to small values of k; (c) zoom of (b) to small
 values of $Re$. }
 \end{figure}

\begin{figure}[htbp]
 \begin{center}
  \includegraphics[width=10cm]{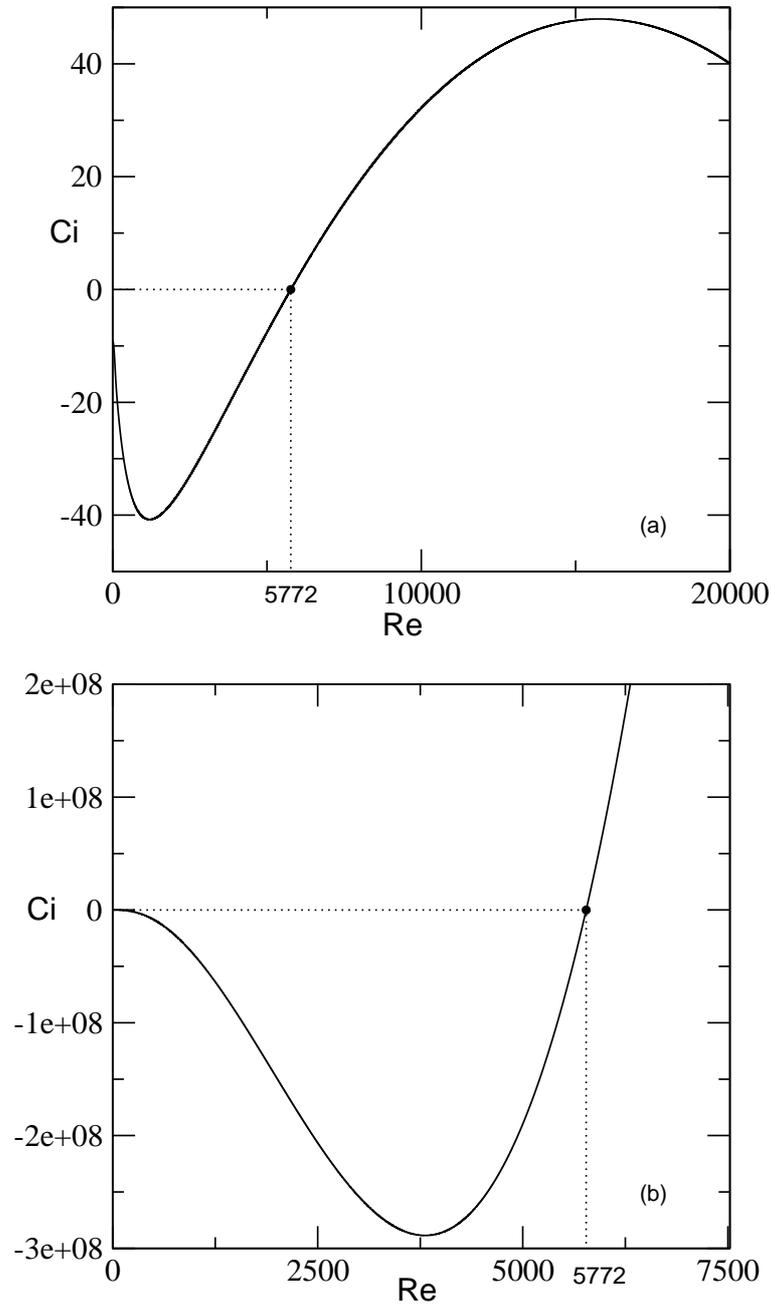}
  \end{center}
  \caption{(a) Growth rate $C_i$ vs. Reynolds number $Re$ for $k=1.02$ in the case of inertial effects normalization showing the 
critical Reynolds number for $2D$-disturbances; (b) Growth rate $C_i$ vs. Reynolds number $Re$ for $k=1.02$ in the case of viscous effects normalization
 showing the critical Reynolds number for $2D$-disturbances. }
 \end{figure}

\newpage
\section{Conclusion}
In this paper, we have investigated the stability of boundary layer in 
Poiseuille flow. We have shown that the instability of the perturbed flow is governed by a remarkably equation named modified
 Orr-Sommerfeld equation.
Contrary to what one might think, we find that Squire's theorem is not applicable for the boundary layer. We find also that normalization
 by inertial or viscous effects leads 
to the same order of stability or instability. For the $2D$ disturbances flow ($\theta=0$), we found the same critical Reynolds 
number for our two normalizations. This value coincides with the one we know for  neutral stability of the known Orr-Sommerfeld 
equation.  We noticed also that for all overs values of $k$ in the case $\theta=0$ correspond the same values of $Re_\delta$ at $c_i=0$ 
whatever the normalization. We therefore conclude that in the boundary layer with a 2D-disturbance, we have the same neutral stability
 curve whatever the normalization. 

\section*{Acknowledgments}
The authors thank IMSP-UAC for financial support.

\end{document}